\documentclass[twocolumn, times]{aastex63}
\usepackage[T1]{fontenc}
\usepackage{amsmath}
\graphicspath{{./}{figures/}}
\usepackage{multirow}

\received{-}
\revised{-}
\accepted{-}
\submitjournal{AAS Journals}

\newcommand{\ms}[1]{$#1~{\rm m\,s^{-1}}$}

\begin{document}

\title{The Excitation Conditions of CN in TW~Hya}

\correspondingauthor{Richard Teague}
\email{richard.d.teague@cfa.harvard.edu}

\author[0000-0003-1534-5186]{Richard Teague}

\affil{Center for Astrophysics | Harvard \& Smithsonian,
       60 Garden Street,
       Cambridge,
       MA 02138,
       USA}
       
\affil{Department of Astronomy,
	   University of Michigan,
       311 West Hall,
       1085 S. University Ave,
       Ann Arbor,
       MI 48109,
       USA}

\author[0000-0002-8932-1219]{Ryan Loomis}

\affil{National Radio Astrophysical Observatory,
	   520 Edgemont Road,
	   Charlottesville,
	   VA 22903,
       USA}
       
\begin{abstract}
We report observations of the cyanide anion, CN, in the disk around TW~Hya covering the $N=1-0$, $N=2-1$ and $N=3-2$ transitions. Using line stacking techniques, 24 hyperfine transitions are detected out of the 30 within the observed frequency ranges. Exploiting the super-spectral resolution from the line stacking method reveals the splitting of hyperfine components previously unresolved by laboratory spectroscopy. All transitions display a similar emission morphology, characterized by an azimuthally symmetric ring, peaking at $\approx 45$~au (0.75\arcsec), and a diffuse outer tail extending out to the disk edge at $\approx 200$~au. Excitation analyses assuming local thermodynamic equilibrium (LTE) yield excitation temperatures in excess of the derived kinetic temperatures based on the local line widths for all fine structure groups, suggesting assumptions of LTE are invalid. Using the 0D radiative transfer code \texttt{RADEX}, we demonstrate that such non-LTE effects may be present when the local H$_2$ density drops to $10^{7}~{\rm cm^{-3}}$ and below. Comparison with models of TW~Hya find similar densities at elevated regions in the disk, typically $z \, / \, r \gtrsim 0.2$, consistent with model predictions where CN is formed via vibrationally excited H$_2$ in the disk atmospheric layers where UV irradiation is less attenuated.
\end{abstract}

\keywords{stuff}

\section{Introduction}
\label{sec:intro}

The formation and subsequent evolution of planetary systems are dictated by the physical structure of their birth environment, the protoplanetary disk. Constraining this structure is therefore of the upmost importance if we are to confront planet formation theories and understand the planet forming potential of the systems we observe.

To this end, excitation analyses of simple molecular species such as CO have proven to be an excellent way to extract gas temperatures at a range of radii and heights within the disk \citep[for example,][]{Schwarz_ea_2016, Zhang_ea_2017}. These works, however, have focused on molecules which, at the large densities expected in disks, are likely in local thermodynamic equilibrium (LTE), therefore offering no information on the density structure. Rather, using a tracer which is no longer in LTE offers the chance for the local collider density, $n({\rm H}_2)$, to be constrained in concert with gas temperature, as done with CS emission in outer 20~au of the disk around TW~Hya \citep{Teague_ea_2018b}.

Due to the large frequency separation of rotational transitions for the simple molecules routinely detected in protoplanetary disks, such excitation analyses require multi-band observations, incurring large uncertainties due to the systematic flux calibration (ranging from a 5\% uncertainty at Band 3 to a 10\% uncertainty at Bands 6 and 7 for ALMA). This can be circumvented by using molecules with (hyper-)fine structure lines, such as CN, HCN, C$_2$H or CH$_3$CN, where multiple transitions are observed in a single frequency tuning \citep{Teague_ea_2016, Hily-Blant_ea_2017, Bergner_ea_2018, Loomis_ea_2018b}. However, with only small shifts in the transition frequencies, many of the (hyper-)fine components can overlap, potentially confusing the interpretations of line ratios.

In this work we consider CN, an exceptionally bright and readily detected molecule with fluxes rivaling that of $^{13}$CO \citep{Guilloteau_ea_2013}. Previously, \citet{Teague_ea_2016} found that for CN in TW~Hya, a simple LTE excitation model of the $N=2-1$ transitions failed outside a radius $\approx 150$~au and inferred surprisingly low excitation temperatures, $T_{\rm ex}$, within this of $\approx 15$~K. With a similar analysis of the $N=3-2$ transition, \citet{Hily-Blant_ea_2017} found comparable low excitation temperatures ranging between 17 and 27~K. More extreme temperatures were found by \citet{Chapillon_ea_2012} who measured excitation temperatures of 8 -- 10~K in the disks of DM~Tau and LkCa~15. 

However, from these results alone it is not possible to distinguish between a scenario where the CN emission arises from cold gas close to the disk midplane and is thermalized, or one where CN emission is sub-thermal and arises from warmer gas in the the low density ($\sim10^6~{\rm cm^{-3}}$) disk atmosphere, as suggested by the modeling of \citet{Cazzoletti_ea_2018}. \citet{Shirley_2015} reports critical densities for the $N=3-2$, $N=2-1$ and $N=1-0$ range between $10^5$ and $10^7~{\rm cm^{-3}}$ at typical disk temperatures of 20 -- 50~K, far exceeding the typical critical densities of CO, $\lesssim 10^4~{\rm cm^{-3}}$, suggesting that collisional excitation may be important for these transitions. The distinction between these two \textbf{excitation} scenarios is not only essential for the interpretation of tests of magnetic field strengths through CN Zeeman-splitting \citep{Vlemmings_ea_2019}, but if CN can be shown to not be in LTE then it may be used effectively as a local gas density tracer.

In this paper we present a combined analysis of three transitions of CN, $N=3-2$, $N=2-1$ and $N=1-0$ in TW~Hya, the closest planet-forming disk at 60.1~pc \citep{Bailer-Jones_ea_2018}. These observations provide the most comprehensive constraints to date on the excitation of CN and subsequently from which regions of the disk the emission is arising. The observations and data reduction are described in Section~\ref{sec:observations}. Both the excitation and kinetic temperatures of the transitions are inferred in Section~\ref{sec:excitation}, with a discussion of the results and a summary following in Sections~\ref{sec:discussion} and \ref{sec:summary}, respectively.

\section{Observations}
\label{sec:observations}

\begin{deluxetable*}{cccDCcDD@{$\;\;\pm$}D}
\tablecaption{Disk Integrated Fluxes of CN\label{tab:observations}}
\tabletypesize{\footnotesize}
\tablehead{
\colhead{$N^{\prime} - N^{\prime\prime}$} &
\colhead{$J^{\prime} - J^{\prime\prime}$} &
\colhead{$F^{\prime} - F^{\prime\prime}$}&
\multicolumn2c{$\nu_0$} &
\colhead{$A_{\rm ul}$} &
\colhead{$g_{\rm u}$}  &
\multicolumn2c{$S_{\rm ij}\mu^2$} &
\multicolumn4c{$\quad$Integrated Flux} \\
\colhead{} &
\colhead{} &
\colhead{} &
\multicolumn2c{(GHz)} &
\colhead{(s$^{-1}$)} &
\colhead{} &
\multicolumn2c{(D)} &
\multicolumn4c{$\quad({\rm Jy\,km\,s^{-1}}$)}
}
\decimals
\startdata
\multirow{5}{*}{1 - 0} & \multirow{5}{*}{3/2 - 1/2} & 3/2 - 1/2 & 113.488120     & 6.74 \times 10^{-6} &  4 &  0.19969 &  0.324 & 0.014     \\
                       &                            & 5/2 - 1/2 & 113.490970     & 1.19 \times 10^{-5} &  6 &  0.62377 &  0.715 & 0.014     \\
                       &                            & 1/2 - 1/2 & 113.499644     & 1.06 \times 10^{-5} &  2 &  0.09661 &  0.258 & 0.014     \\
                       &                            & 3/2 - 3/2 & 113.508907     & 5.19 \times 10^{-6} &  4 &  0.08623 &  0.267 & 0.014     \\
                       &                            & 1/2 - 3/2 & 113.520432     & 1.30 \times 10^{-6} &  2 & -0.81635 &  0.026 & 0.014     \\[3pt]
\hline
\multirow{8}{*}{2 - 1} & \multirow{3}{*}{3/2 - 1/2} & 5/2 - 3/2 & 226.659558     & 9.47 \times 10^{-5} &  6 &  0.62232 &  2.830 & 0.058     \\
                       &                            & 1/2 - 1/2 & 226.663693     & 8.46 \times 10^{-5} &  2 &  0.09660 &  1.033 & 0.058     \\
                       &                            & 3/2 - 1/2 & 226.679311     & 5.27 \times 10^{-5} &  4 &  0.19154 &  1.211 & 0.058     \\[3pt] \cline{2-13}
                       & \multirow{5}{*}{5/2 - 3/2} & 5/2 - 3/2 & 226.874191     & 9.62 \times 10^{-5} &  6 &  0.62813 &  2.829 & 0.051^{b} \\
                       &                            & 7/2 - 5/2 & 226.874781     & 1.14 \times 10^{-4} &  8 &  0.82793 &  3.929 & 0.051^{b} \\
                       &                            & 3/2 - 1/2 & 226.875896     & 8.59 \times 10^{-5} &  4 &  0.40262 &  2.014 & 0.051^{b} \\
                       &                            & 3/2 - 3/2 & 226.887420     & 2.73 \times 10^{-5} &  4 & -0.09492 &  0.637 & 0.051     \\
                       &                            & 5/2 - 5/2 & 226.892128     & 1.81 \times 10^{-5} &  6 & -0.09744 &  0.646 & 0.051     \\[3pt]
\hline
\multirow{18}{*}{3 - 2}& \multirow{7}{*}{5/2 - 5/2} & 3/2 - 3/2 & 339.446777     & 2.26 \times 10^{-5} &  4 & -0.7013  &  0.115 & 0.058^{c} \\
                       &                            & 3/2 - 5/2 & 339.459996     & 4.33 \times 10^{-6} &  4 & -1.4197  & -0.005 & 0.080     \\
                       &                            & 5/2 - 3/2 & 339.462635     & 2.95 \times 10^{-6} &  6 & -1.4104  &  0.019 & 0.080     \\
                       &                            & 5/2 - 5/2 & 339.475904     & 2.12 \times 10^{-5} &  6 & -0.55307 &  0.476 & 0.080     \\
                       &                            & 5/2 - 7/2 & 339.493211     & 2.99 \times 10^{-6} &  6 & -1.4051  &  0.061 & 0.080     \\
                       &                            & 7/2 - 5/2 & 339.499288     & 2.33 \times 10^{-6} &  8 & -1.3872  &  0.073 & 0.080     \\[3pt] \cline{2-13}
                       & \multirow{6}{*}{5/2 - 3/2} & 3/2 - 5/2 & 339.992257     & 3.89 \times 10^{-6} &  4 & -1.4679  &  0.004 & 0.082     \\
                       &                            & 5/2 - 5/2 & 340.008126     & 6.20 \times 10^{-5} &  6 & -0.09005 &  1.474 & 0.082     \\
                       &                            & 3/2 - 3/2 & 340.019626     & 9.27 \times 10^{-5} &  4 & -0.09129 &  1.448 & 0.082     \\
                       &                            & 7/2 - 5/2 & 340.031549     & 3.85 \times 10^{-4} &  8 &  0.82749 &  8.743 & 0.082     \\
                       &                            & 3/2 - 1/2 & 340.035269^{a} & 2.89 \times 10^{-4} &  4 &  0.40197 &  9.914 & 0.082^{b} \\
                       &                            & 5/2 - 3/2 & 340.035507^{a} & 3.23 \times 10^{-4} &  6 &  0.62698 &  9.914 & 0.082^{b} \\[3pt] \cline{2-13}
                       & \multirow{5}{*}{7/2 - 5/2} & 7/2 - 5/2 & 340.247590^{a} & 3.93 \times 10^{-4} &  8 &  0.82118 & 16.991 & 0.083^{b} \\
                       &                            & 9/2 - 7/2 & 340.247861^{a} & 4.13 \times 10^{-4} & 10 &  0.95478 & 16.991 & 0.083^{b} \\
                       &                            & 5/2 - 3/2 & 340.248544     & 3.67 \times 10^{-4} &  6 &  0.68198 &  8.461 & 0.083^{b} \\
                       &                            & 5/2 - 5/2 & 340.261773     & 4.48 \times 10^{-5} &  6 & -0.2321  &  1.070 & 0.083     \\
                       &                            & 7/2 - 7/2 & 340.264949     & 3.35 \times 10^{-5} &  8 & -0.2333  &  1.104 & 0.083     \\[3pt]
\enddata
\tablecomments{The line frequencies, Einstein~A coefficients, upper state degeneracies and dipole strengths were taken from the CDMS \citep{CDMS}, collating the results of \citet{Dixon_Woods_1977} and \citet{Skatrud_ea_1983}.}
\tablenotetext{^{\rm a}}{Updated frequency based on Gaussian fits to the spectra.}
\tablenotetext{^{\rm b}}{Either fully or partially blended transitions.}
\tablenotetext{^{\rm c}}{Transition lies at edge of spectral window and is thus only a lower limit to the integrated flux.}
\end{deluxetable*}

We combine ALMA Band 3, 6 and 7 observations spanning the the $N=1-0$, $N=2-1$ and $N=3-2$ transitions at 113.5~GHz, 226.8~GHz and 340.1~GHz, respectively. The Band 6 data (2013.1.00387.S, PI S. Guilloteau) were previously published in \citet{Teague_ea_2016}, however the Band 3 data (2017.1.01199.S, PI R. Loomis) and Band 7 (2016.1.00440.S, PI R. Teague) have not been previously published. In the following sections we briefly describe the data reduction and imaging process for each set of observations.

\subsection{Calibration}

\subsubsection{Band 3}

Band 3 observations, part of the Cycle 5 project 2017.1.01199.S (PI: R. Loomis), were taken the two consecutive nights, September 15 and 16, 2018. Each execution, 61 minutes each, used 42 antennas spanning baselines from 15~m to 1261~m, and included 39 minutes of on-source integration, totaling 78 minutes on source. J1107-4449 was used for all three executions as bandpass and flux calibrator, while J1037-2934 was used for phase calibration. The correlator was tuned to cover the $N = 1-0$, $J = 3/2 - 1/2$ components around 113.491~GHz at a 31~kHz ($80~{\rm m\,s^{-1}}$) spectral resolution. The data were iteratively phase self-calibrated for three rounds with a final solution interval of 30~s.

\subsubsection{Band 6}

Observations were performed on May 13, 2015 as part of the Cycle 2 observations for project 2013.1.00387.S (PI: S. Guilloteau), using 34 antennae and integrated on source for 46 minutes. The correlator was set up to cover the $N=2-1$, $J=3/2-1/2$ and $J=5/2-3/2$ fine structure groups at a channel spacing of 30~kHz ($40~{\rm m\,s^{-1}}$). With baselines only extending to 550~m, these observations were the most compact of the three. The data were phase self-calibrated using a solution interval of 30~s.

\subsubsection{Band 7}

These observations, using 45 antennas, were taken December 2, 2016 as part of a Cycle 4 project 2016.1.00440.S (PI: R. Teague). Baselines ranged between 15 and 700~m, with a total on-source time of 48 minutes. The correlator was set up to cover the $J=5/2-5/2$, $J=5/2-3/2$ and $J=7/2-5/2$ fine structure groups with a channel spacing of 244~kHz ($215~{\rm m\,s}^{-1}$). Data were self-calibrated using a solution interval of 30~s.

\subsection{Imaging}

All imaging was performed with the \texttt{tclean} function with a multi-scale approach and Briggs weighting. We adopted a Keplerian mask\footnote{The code to make Keplerian masks is available from \url{https://github.com/richteague/keplerian_mask}} for the \texttt{CLEAN} mask, which was generated assuming a source inclination of 6.8\degr{}, a position angle of 151\degr{}, a stellar mass of 0.65~$M_{\rm sun}$ and a distance of 60.1~pc. A convolution with a Gaussian kernel with a FWHM of 1\arcsec{} was used to ensure that all emission was captured within the \texttt{CLEAN} mask. For each Band, a dirty image was produced in order to measure the RMS outside the Keplerian mask in order to define the threshold for the subsequent \texttt{CLEAN}ing, which was set to twice this value.

We adopted a different robust values for each Band in order to yield the best trade-off between spatial resolution and sensitivity, while yielding comparable synthesized beams. For Band 3, a robust value of 0 was chosen yielding a beam of $0.54\arcsec \times 0.48\arcsec$ at a position angle of 46.1\degr{} and an RMS of $2.7~{\rm mJy~beam^{-1}}$ in a line free channel. For Band 6, a robust value of -1 was adopted, resulting in a beam of $0.52\arcsec \times 0.43\arcsec$ with a position angle of 92.2\degr{} and an RMS of $5.2~{\rm mJy~beam^{-1}}$ in a line-free channel. Finally, a robust value of 0.5 was chosen for Band 7, resulting in a beam of $0.31\arcsec \times 0.29\arcsec$ at a position angle of 61.5\degr{}, and an RMS of $3.6~{\rm mJy~beam^{-1}}$ meausred in a line free channel.

\subsection{Observational Results}
\label{sec:observations:results}

\begin{figure*}
\centering
\includegraphics[width=\textwidth]{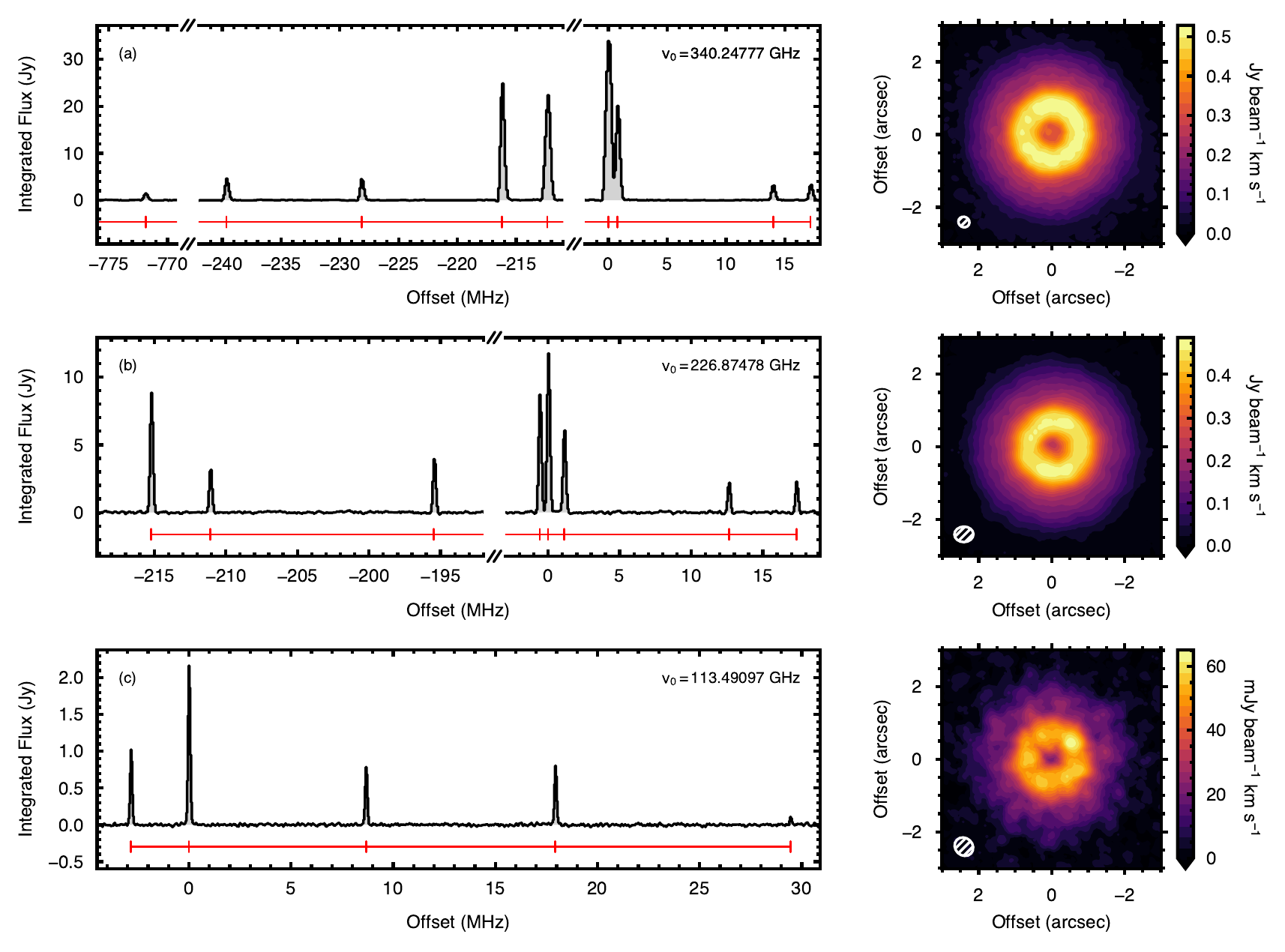}
\caption{The spatially integrated intensities (left column) and spectrally integrated flux densities (right column) are shown for the $N=3-2$, $N=2-1$ and $N=1-0$ lines, top to bottom. The integrated spectra were calculated using the shifting and stacking method described below. Locations of hyperfine components are denoted by the red ticks. The zeroth moment maps, created with \texttt{bettermoments}, were masked with the CLEAN mask and clipped below $2\sigma$ where $\sigma = \{3.6,\, 5.2,\, 3.5\}~{\rm mJy~beam^{-1}}$, for $N=3-2$, $N=2-1$ and $N=1-0$, respectively. \label{fig:observations}}
\end{figure*}

To boost the signal to noise of the spectra, we follow the line shifting and stacking technique described in \citet{Teague_ea_2016} and \citet{Yen_ea_2016}, implemented in the Python package \texttt{GoFish} \citep{GoFish}. To verify the detection of weak components with the image stacking, we used a matched-filter approach implemented in \texttt{VISIBLE}\footnote{\url{https://github.com/AstroChem/VISIBLE}} \citep{Loomis_ea_2018b}, using the well-detected components as a filter.

To infer the velocity profile used to deproject the image cubes, we first split the data into concentric annuli with a width one quarter of the beam major FWHM, assuming an inclination of $6.5\degr$ and a position angle of $151\degr$. Then, for each annulus we follow the method described in \citet{Teague_ea_2018c} to infer the projected rotational velocity. As this method uses a Gaussian Process to model the aligned and stacked spectra, it is capable of handling spectra with multiple overlapping components, as is the case for CN hyperfine transitions. We obtain a projected rotation curve for each fine-structure group, which are then fit with a Keplerian rotation profile. All profiles were consistent with a stellar mass of $M_{\rm star} = 0.66 \pm 0.02~M_{\rm sun}$, where the uncertainty is the standard deviations of the measurements. This is consistent with previously determined $\sqrt{M_{\rm star}} \sin(i)$ measurements \citep[e.g.][]{Teague_ea_2016, Huang_ea_2018}.

To measure the disk integrated fluxes for the individual components, the spectra were aligned to a common line center assuming the inferred Keplerian rotation profile and stacked, binning down to a velocity resolution of $40~{\rm m\,s^{-1}}$. For each transition the stacked spectrum was then integrated over a window with a width of $1.5~{\rm km\,s^{-1}}$, centered on the line center. The integrated fluxes over each annulus were summed for the final results shown in Table~\ref{tab:observations} where the uncertainties were calculated following Eqn.~16 from \citet{Yen_ea_2016}.

Figure~\ref{fig:observations} shows an overview of the observations with integrated spectra, including the aligning and stacking described above, in the left column and zeroth moment (or integrated flux density maps) on the right. The aligning and stacking of the data removes the typical double-peaked profile characteristic of rotating sources. We note that missing short-spacings for the Band 7 observations (but also the Band 6 data to some extent) result in regions of negative flux in restored images. This only affects the channels close to the systemic velocity of the system with the emission extends the full diameter of the disk ($\approx 8\arcsec$), as opposed to the edge channels where emission is confined to much smaller spatial scales. For the integrated spectra in Fig.~\ref{fig:observations}, we use an outer radius of 4\arcsec{} for all three bands to ease comparison between bands.

\begin{figure}
    \centering
    \includegraphics{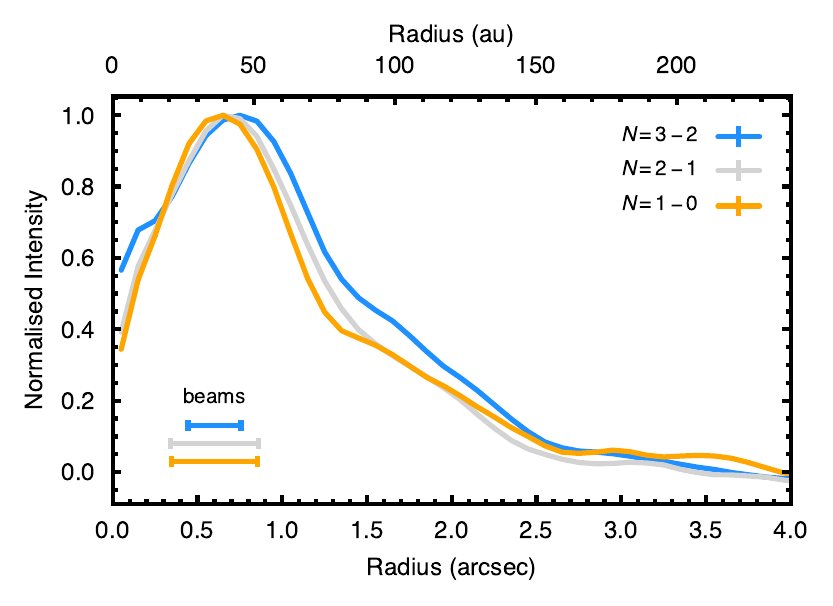}
    \caption{Radial intensity profiles for the three collapsed transitions of CN shown in Fig.~\ref{fig:observations}. Error bars showing 16th to 84th percentile of each radial bin are also included, however are unable to be easily seen as they are a similar size to the line widths. The beam sizes are shown in the bottom left of the figure.}
    \label{fig:radial_profiles}
\end{figure}

Figure~\ref{fig:radial_profiles} shows the radial profile of the shifted and stacked spectra using the \texttt{radial\_profile} function in \texttt{GoFish}. The advantage of using this rather than an azimuthally averaged zeroth moment (total intensity) map is that this is much more sensitive to the inner and outer regions of the disk which would typically be lost when using a $\sigma$ cut on the data to make the moment map \citep[e.g.][]{Loomis_ea_2015}. All three transitions appear to have the same morphology: a broad ring peaking at 0.7\arcsec{} (42~au), with a broader excess out to $\approx 3\arcsec$ (180~au) and a knee feature at $\approx 1.5\arcsec$ (90~au), most prominent in the $N-1-0$ components, however visible to some extent with the higher energy transitions. The $N=3-2$ emission lines appear to have an increase in the inner disk relative to the other lower energy transitions. When convolving the images down to the same spatial resolution, this enhancement in the inner disk persists, suggesting that it is a real increase in flux.

Such radial profiles are comparable to that observed in both C$_2$H and CS \citep{Bergin_ea_2016, Teague_ea_2018b}, and was predicted by \citet{Cazzoletti_ea_2018} as a natural outcome due to UV-modulated formation via vibrationally excited H$_2$. Interestingly, the peak in CN rings lies radially inwards of the C$_2$H rings which peak at 1.1\arcsec{} \citep{Bergin_ea_2016}. For all three transitions we observe a depletion of emission at the disk center, also predicted by the models in \citet{Cazzoletti_ea_2018}.

\subsection{Blended Components}
\label{sec:observations:blended_components}

\begin{figure*}
    \centering
    \includegraphics[]{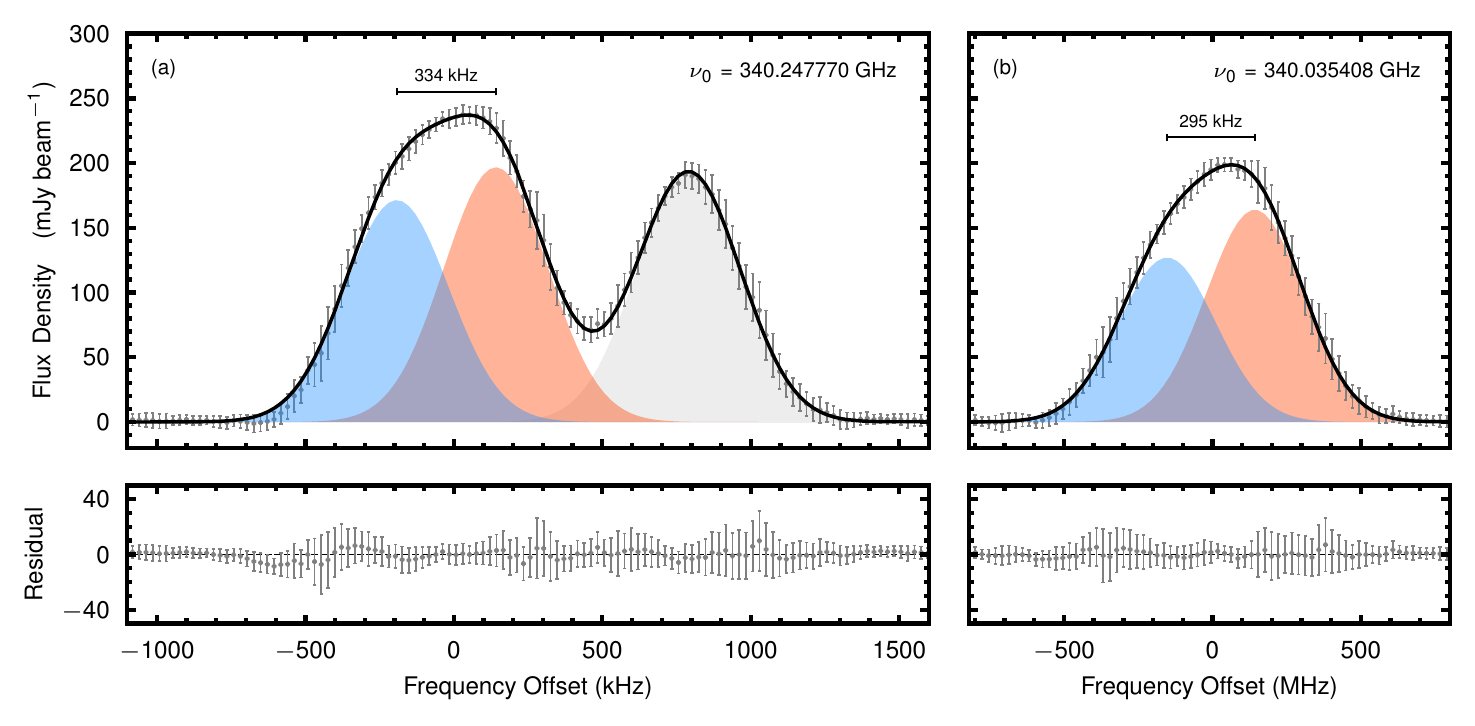}
    \caption{Demonstrating the power of super-sampled spectra in distinguishing blended components. The top panels show deprojected spectra re-sampled at \ms{20} with the small gray error bars. Clear departure from a single Gaussian component is seen, althought the morphology well fit by two Gaussian components shown in blue and red, with the black line showing the combined spectrum. The residuals are shown in the bottom panels.}
    \label{fig:blended_hyperfines}
\end{figure*}

By resampling the shifted and stacked spectrum we are able to better resolve the line profile. While this resampling does not remove any systematic effects, such as the impact of the spectral response function or smearing due to the finite spatial resolution of the data, it does allow for a clearer characterization of such features.

Two components of the $N = 3-2$ transition at 340.035408~GHz and 340.247770~GHz show a distinctive slanted top, contrary to the Gaussian-shaped profiles for the other transitions. Given the low spectral resolution of the $N = 3 - 2$ data, these profiles are unable to be seen in individual pixels, however comparison with higher spectral resolution data published in \citet{Hily-Blant_ea_2017} (their Fig.~C.3) confirms the presence of this feature without the line stacking.

These components are known to be blends with their line centers current reported as identical, albeit with relatively large uncertainties of 50~kHz \citep{CDMS}. As shown in Fig.~\ref{fig:blended_hyperfines}, these line profile can be well described as the sum of two Gaussian components. Making the assumption that blending between the components is negligible, that is, the lines can be modeled as the simple sum of two components, we fit the profile with two offset Gaussians which sharing the same line width but independent amplitudes to measure possible offsets in the line centers. For the fitting of the 340.247770~GHz pair of components, we include a third component at 340.248544~GHz which shares the same line width.

The resulting line centers are quoted in Table~\ref{tab:observations} with a precision of $\sim 5$~kHz. We have assigned the line center of the weaker of the two Gaussians to the blended component with the lower $S_{ij}\mu^2$ value. The split between the lines, 334~kHz for the 340.247770~GHz components and 295~kHz for the 340.035408~GHz components, are larger than laboratory uncertainties, but the true level of splitting is likely limited by both our assumption of no blending and the 244~kHz resolution of the observations, which will broaden the lines somewhat. High spectral resolution observations are required to confirm the quoted offsets, either astronomical or laboratory spectroscopic measurements.

\section{Inferring the Excitation Conditions}
\label{sec:excitation}

For a molecular line in full local thermodynamic equilibrium, LTE, the excitation temperature of the line is equal to the source region kinetic temperature, $T_{\rm ex} = T_{\rm kin}$. If this equality does not hold then either sub-thermal or super-thermal excitation is taking place. In this section we measure both the excitation temperature and kinetic temperature of the five fine-structure groups (groups of transitions which share the same $N$ and $J$ quantum numbers) to explore the excitation conditions of CN in TW Hya.

\begin{figure*}
    \centering
    \includegraphics[width=\textwidth]{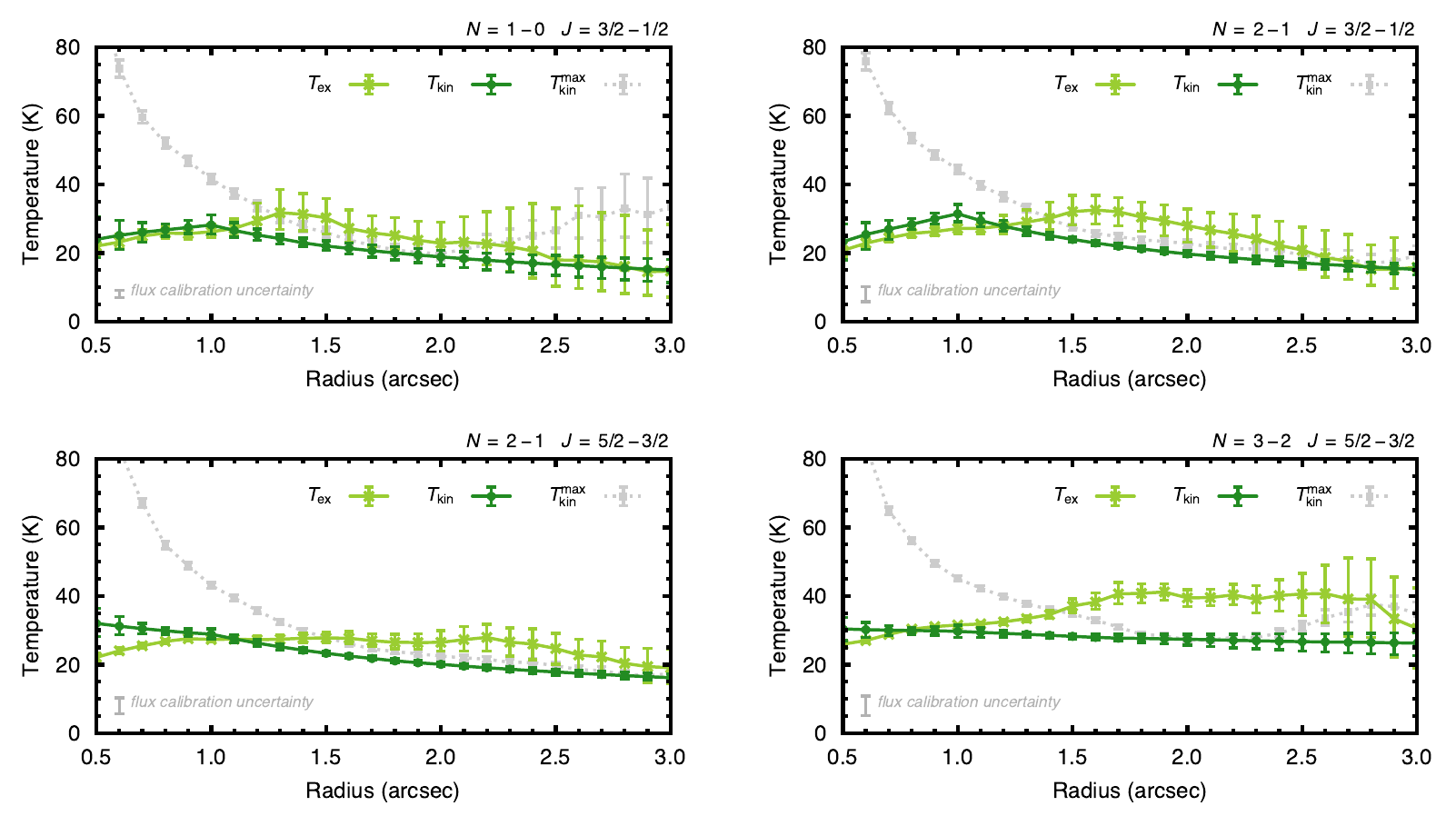}
    \caption{Comparisons of the $T_{\rm ex}$ profiles in light green circles, for four of the fine structure group with the inferred $T_{\rm kin}$ profiles show by dark green squares. The gray squares shows $T_{\rm kin}^{\rm max}$, the maximum kinetic temperature consistent with the measured line widths. The error bars represent the 16th to 84th percentile of 350 random draws from the posterior distributions. The typical systematic uncertainty on $T_{\rm ex}$ associated with the flux calibration is shown in the lower left of each panel presenting $5\%$ for the $N = 1-0$ transition and $10\%$ for the $N=2-1$ and $N=3-2$ transitions.}
    \label{fig:TexTkin}
\end{figure*}

\subsection{Excitation Temperature}
\label{sec:excitation:Tex}

The excitation temperatures are inferred by modeling the spectra, following the method described in \citet{Hily-Blant_ea_2013}, discussed thoroughly in Appendix~\ref{sec:app:Tex}. In brief, by assuming that all transitions in a fine-structure group share the same $T_{\rm ex}$, then the total optical depth of that fine-structure group, $\tau_0$, can be distributed between each component based on the relative intensity of that component. Then, assuming that each component has a Gaussian line profile with a shared line width, $\Delta V$, it is possible to model the emergent intensity from an isothermal slab with only $\{T_{\rm ex},\, \Delta V,\, \tau_0\}$. We note that unlike the method used in \citet{Hily-Blant_ea_2017}, we do not take the Rayleigh-Jeans approximation but rather calculate the full Planck function.

We apply this method only to four of the six fine structure groups. The $N = 3-2$ $J=5/2-5/2$ and $J=7/2-5/2$ were unable to be fit as the former only had one significantly detected transition, while the latter had a complex triple blend which was unable to be well reproduced with our modeling approach.

Each image cube covering a fine-structure group was radially binned into 0.1\arcsec{} wide annuli, then spectrally deprojected and stacked to produce an average spectrum at each radius, as described in Section~\ref{sec:observations:results}. The Affine-invariant MCMC ensemble sampler \texttt{emcee} was used to explore the posterior distributions of $\{T_{\rm ex},\, \Delta V,\, \tau_0\}$ for each averaged spectrum. A thorough description of this procedure is described in Section~\ref{sec:app:Tex}. We note that the flux calibration uncertainties of 5\% in Band 3 and 10\% for Band 6 and 7 result in comparable uncertainties in $T_{\rm ex}$ and $\tau_0$, but $\Delta V$ is unaffected. The radial profiles of $T_{\rm ex}$ are shown as red-squares in Fig.~\ref{fig:TexTkin}.

In addition to the full spectrum fits, we additionally fit for the integrated fluxes and peak flux density. This removes the dependence on the line profile as a check that the spectral shifting and stacking method in Section~\ref{sec:observations:results} does not influence the fit results. As many of the components are blended, we break each fine structure group into the smallest non-overlapping sections to calculate the integrated flux. The peak flux density is measured as the peak value in each of these sections. We find no difference between these results and the full spectrum fits, suggesting that the deprojection does not affect our modeling approach.

\subsection{Kinetic Temperature}
\label{sec:excitation:Tkin}

As the emission lines in a protoplanetary disk are dominated by Doppler broadening, the line width is directly proportional to $\Delta V$,

\begin{equation}
    \Delta V = \sqrt{\frac{2kT_{\rm kin}}{\mu_{\rm CN}m_p} + v_{\rm turb}^2},
\end{equation}

\noindent where $k$ is Boltzmann's constant, $\mu_{\rm CN} = 26$ and is the molecular weight of CN, $m_p$ is the proton mass and $v_{\rm turb}$ is the turbulent velocity dispersion. Previous attempts to measure $v_{\rm turb}$ in TW~Hya have shown that this is a negligible component, corresponding to an uncertainty on $T_{\rm kin}$ on the order of $\sim 1$~K \citep{Hughes_ea_2011, Teague_ea_2016, Teague_ea_2018b, Flaherty_ea_2018}. A larger effect is the broadening due to the finite beam size of the observations, as described in \citet{Teague_ea_2016}. We undertake an extensive forward modeling process, thoroughly described in Appendix~\ref{sec:app:Tkin}, to estimate the impact of this broadening and, based on the measured $\Delta V$ profiles from Section~\ref{sec:excitation:Tex}, infer the underlying $T_{\rm kin}$ profile for all fine-structure groups. Unlike for the measurement of $T_{\rm ex}$, the line width is unaffected by flux calibration uncertainties. Both single and double power law profiles of the underlying $T_{\rm kin}$ profile were tested and gave comparable results. Both were characterized by a very shallow exponent, similar to that found in \citet{Teague_ea_2016}. The results of the posterior distribution exploration are described in Table~\ref{tab:tkin}.

\subsection{Results}
\label{sec:excitation:results}

The inferred $T_{\rm ex}$ and $T_{\rm kin}$ profiles are shown in Fig.~\ref{fig:TexTkin}. Between 0.5\arcsec{} and 3\arcsec{} (30~au and 180~au) the average excitation temperature is $\langle T_{\rm ex} \rangle = 27 \pm 7$~K, while the average kinetic temperature is $\langle T_{\rm kin} \rangle = 24 \pm 5$~K, where the uncertainties describe the standard deviation of the parameters. These values are broadly consistent with each other, suggesting a thermalization of the transitions, as found in \citet{Hily-Blant_ea_2017}. We do find elevated values for $T_{\rm ex}$ in the outer disk compared to both \citet{Teague_ea_2016} and \citet{Hily-Blant_ea_2017} for the $N = 2-1$ and $N=3-2$ transitions, respectively. This is due to this work not using the Rayleigh-Jeans approximation of the Planck function, as discussed in Appendix~\ref{sec:app:Tex}.

Although the disk average kinetic and excitation temperatures are broadly consistent, we do find that over large radial extents we infer $T_{\rm ex}$ values which exceed $T_{\rm kin}$, most notably for the higher energy $N = 3-2$ transition. A similar result was reported in \citet{Teague_ea_2016} using the same $N=2-1$ data presented here. Typically $T_{\rm ex} > T_{\rm kin}$ suggests super-thermal excitation arising from non-LTE effects and these results are discussed in Section~\ref{sec:discussion:nonLTE}.

Unlike the excitation temperature, no assumptions need to be made on the thermalization of the transitions for the calculation of the the kinetic temperature, therefore providing a more robust upper limit to gas temperature. The kinetic temperatures suggest that CN is tracing a moderately warm region in the disk around 20 -- 30~K. This is comparable in temperature to the gas probed by CS and CH$_3$CN emission in TW~Hya \citep{Teague_ea_2018b, Loomis_ea_2018b}. While non-thermal broadening components, such as unresolved turbulence, can result in a lower kinetic temperature, previous searches for such broadening in TW~Hya suggest that any such component would be negligible \citep{Teague_ea_2016,  Teague_ea_2018b, Flaherty_ea_2018}. Therefore for the CN emission observed in TW~Hya, it is unlikely any of it is tracing regions with temperatures $T < 20$~K.

\begin{figure}
    \centering
    \includegraphics[width=\columnwidth]{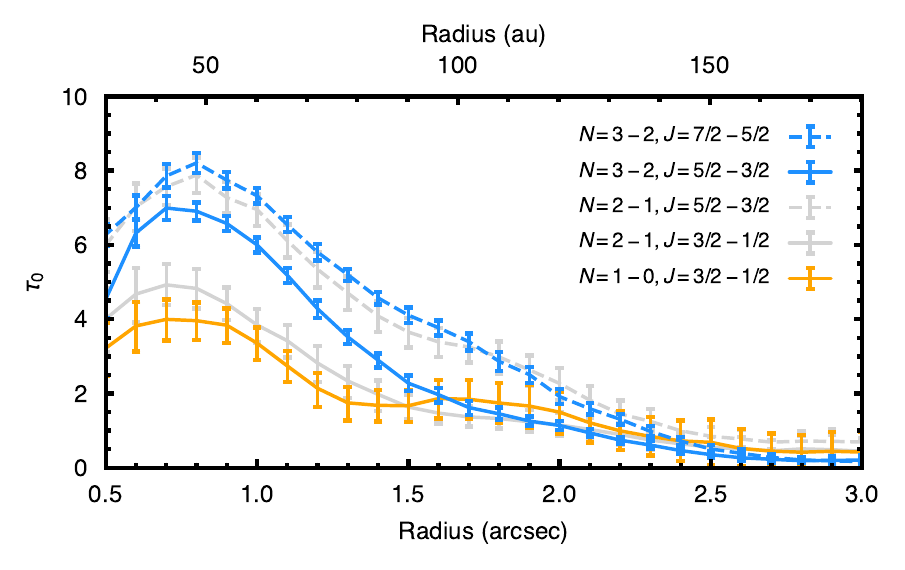}
    \caption{Inferred fine-structure optical depths, $\tau_0$ from the excitation analysis. The error bars show the 16th to 84th percentiles of the posterior distributions. Note that each fine structure group consists of several hyperfine components among which this optical depth is shared, such that the optical depth of an individual component is $\lesssim 30\%$ of this value.}
    \label{fig:optical_depth}
\end{figure}

We also show the inferred optical depths (the total for each fine-structure group; Eqn.~\ref{eq:app:tau}) in Fig.~\ref{fig:optical_depth}, finding comparable values to those in \citet{Hily-Blant_ea_2017}. These trace a similar profile as the radial intensity profiles shown in Fig.~\ref{fig:radial_profiles}, suggesting that changes in the column density are responsible for the radial morphology rather than changes in the gas temperature. The `knee' feature at $\approx 1.5\arcsec$ is most pronounced in the $N = 1-0$ emission, while the higher optical depth fine-structure groups appear to show less of a feature.

\section{Discussion}
\label{sec:discussion}

\subsection{Non-LTE Excitation}
\label{sec:discussion:nonLTE}

We find evidence for super-thermal excitation in the outer disk for the $N = 2-1$ and $N = 3-2$ transitions, similar to the previous analysis described in \citet{Teague_ea_2016}. We note that \citet{Hily-Blant_ea_2017} do not find such evidence, but we believe there is a mistake in their calculation of $T_{\rm kin}$ which misses a factor of $\sqrt{2}$ due to $\Delta V$ being the Doppler width, $\Delta V = \sqrt{2}\sigma_{v}$, in the calculation of the thermal line width (Eqn.~\ref{eq:app:dV}). After taking this factor into account, we find similar evidence for super-thermal excitation. At face value, this suggests non-LTE effects are at play with certain hyperfine components being radiatively pumped. However, a more likely scenario is that the assumptions made in the modeling of the spectra are not applicable.

One assumption is that the optical depth of various hyperfine components can be treated independently and thus a total optical depth can be modeled as the sum of the optical depths of the components (see Eqn.~\ref{eq:app:tau}). As demonstrated in Fig.~\ref{fig:observations}, there are multiple components within both these transitions which can overlap, particularly in the inner disk where the intrinsic line widths are large. Further to this, the modeling of the spectra resulted in large optical depths, $\tau \gtrsim 1$, for many of the components. In this regime such blended components may causes issues and a radiative transfer code which takes into account such blending would be requires, see the discussion in Sect.~\ref{sec:discussion:future_direction}.

\begin{figure}
    \centering
    \includegraphics[width=\columnwidth]{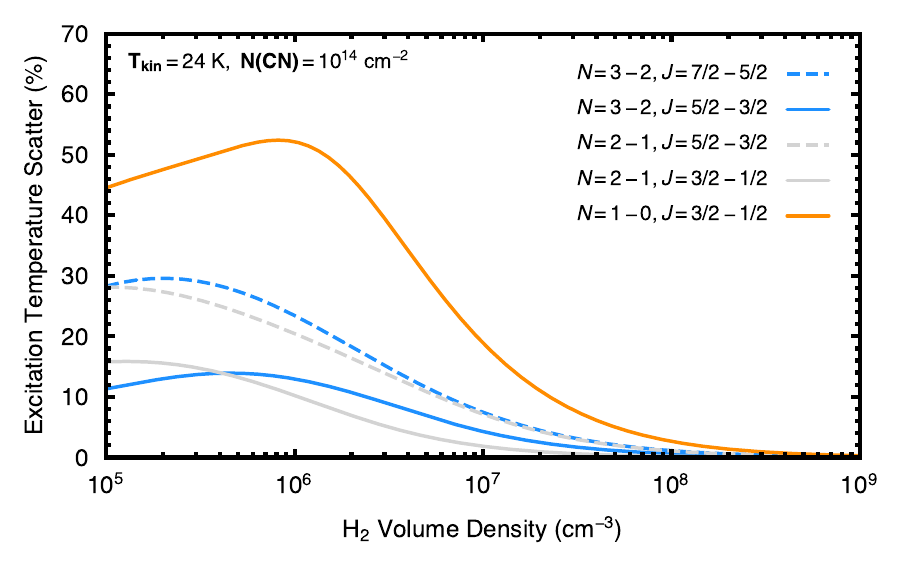}
    \caption{Showing the standard deviation of the $T_{\rm ex}$ values for each fine-structure group as a function of H$_2$ collider density. The kinetic temperature is fixed at $T_{\rm kin} = 24$~K and we assume a CN column density of $N({\rm CN}) = 10^{14}~{\rm cm^{-2}}$. The linewidth is assumed to be purely thermal.}
    \label{fig:Tex_scatter}
\end{figure}

A second assumption made is that the hyperfine components within a fine structure group share an excitation temperature. To test this assumption we use the 0-D non-LTE radiative transfer code \texttt{RADEX} \citep{vanderTak_ea_2007} with the latest CN collisional rates which include hyperfine structure \citep{Kalugina_Lique_2015}, obtained from the LAMDA database\footnote{\url{https://home.strw.leidenuniv.nl/~moldata/}} \citep{Schoeler_ea_2005}. Assuming a $T_{\rm kin} = 24~{\rm K}$, the average $T_{\rm kin}$ from Section~\ref{sec:excitation:Tkin}, a fully thermalised linewidth. and a CN column density of $N({\rm CN}) = 10^{14}~{\rm cm^{-2}}$, we calculate the excitation temperatures for a range of H$_2$ collider densities assuming a thermal H$_2$ ortho-para distribution. The results are shown in Fig.~\ref{fig:Tex_scatter}.

Each line shows one of the fine structure groups described in Table~\ref{tab:observations}. The y-axis measures the scatter in the $T_{\rm ex}$ values measured within a fine-structure group. For an entirely thermalised fine-structure group, this should be close to zero as all components share the same excitation temperature. Even at collider densities as high as $n_{\rm H_2} \sim 10^7~{\rm cm^{-3}}$ deviations of up to 10\% are found within a fine-structure group, with the lower energy $N=1-0$ transition being the most sensitive. Changing the physical properties of this model gives quantitatively different curves, however the same overall trend: deviations in $T_{\rm ex}$ across fine-structure groups are found for $n_{\rm H_2} \lesssim 10^7~{\rm cm^{-3}}$.

This analysis raises two points. Firstly, it suggests that the elevated $T_{\rm ex}$ values found in Section~\ref{sec:excitation:Tex} could be explained by slight non-LTE effects which could manifest if the CN emission arises from regions of the disk where $n_{\rm H_2} \lesssim 10^7~{\rm cm^{-3}}$. Based on the physical structure model from \citet{Qi_ea_2013}, this is achieved at heights of $z\,/\,r \gtrsim 0.2$. In the model, this vertical region has a gas temperature of 20 -- 30~K, again consistent with the average $T_{\rm kin}$ values found. The implications of this on the formation of CN are discussed later.

Secondly, the \texttt{RADEX} analysis suggested that the $N = 1-0$ transition should be more sensitive than the higher energy transitions, counter to what is found in Section~\ref{sec:excitation:results}. A likely explanation is that the $N=1-0$ transition arises from a slightly deeper region than $N=2-1$ and $N=3-2$ transitions, as routinely observed with other molecular species, e.g. CO \citep{Dartois_ea_2003}. Such layers should be readily distinguished by the thermal widths of the lines, however the spatial and spectral resolution of the data to hand preclude us from doing so here.

This difference in sensitivity to the local collider density could also potentially explain why the `knee' feature at $\approx 1.5\arcsec$ is more pronounced for the lower energy $N = 1-0$ transition. A large depletion in the gas surface density is known to exist at $\approx 1.5\arcsec$ from observations of scattered light \citep[e.g.][]{vanBoekel_ea_2017} and molecular line emission \citep[e.g][]{Teague_ea_2017}. As a depletion in the gas surface density would cause a comparable drop in the local collider (H$_2$) density, a more significant change in the excitation of the $N=1-0$ transition would be seen than for the two higher energy transitions.

\subsection{Formation Pathway of CN}
\label{sec:discussion:formation}

\citet{Cazzoletti_ea_2018} demonstrated that the annular morphology of the CN emission in TW~Hya, as well as in other disks, can be readily replicated when CN is formed primarily through a pathway involving vibrationally excited H$_2$. The authors argued that the upper atmospheric layers of the disk, bathed in higher fluxes of stellar FUV, are more conducive to the formation of CN owing to the larger abundance of H$_2^{*}$. \citet{Cazzoletti_ea_2018} found in their models that CN was primarily formed at heights of $z\,/\,r \sim 0.4$, considerably higher than suggested by the inferred $T_{\rm kin}$ values in Section~\ref{sec:excitation:Tkin}. However, the precise relation between the UV flux and a $z\,/\,r$ location will be dependent on the flaring of the disk which dictates how readily stellar radiation is deposited at a given location in the disk. A source-specific model, in which the flaring angle has been constrained through independent observations, would be necessary to make a robust comparison for the formation environment of the CN. 

More recently, \citet{Arulanantham_ea_2020} used Hubble Space Telescope observations of a sample of disks in the Lupus star forming region to relate the FUV fluxes of the stars with the CN emission previously reported by \citet{vanTerwisga_ea_2019}. The authors found a negative correlation between the integrated CN $N = 3-2$ flux and the FUV continuum and C~{\sc ii}$]$ fluxes which they attributed to efficient dissociation of CN in the FUV-dominated upper layers. Such a scenario could explain why CN appears to arise from deeper regions in TW~Hya as although the elevated FUV-dominated layers have a higher formation rate, they also have a higher destruction rate through dissociation and thus deeper layers are more favorable for larger steady-state abundances.

In sum, these results suggest that the CN emission observed in TW~Hya arise from moderately elevated regions tracing regions with gas temperatures of 20 -- 30~K, consistent with the UV-driven formation pathway described in \citet{Cazzoletti_ea_2018}. We interpret the low excitation temperatures reported in \citet{Teague_ea_2016} and \citet{Hily-Blant_ea_2017} as due to non-LTE excitation effects rather than gas temperatures lower than 20~K. Ultimately, observations of inclined disks are necessary to truly reveal at what altitude CN is most abundant and thus distinguish between formation mechanisms. Edge on sources such as the Flying Saucer \citep{Dutrey_ea_2017} provide the best viewing angle, however with an increasing number of methods used to infer the emission height from high spatial resolution \citep[e.g.][]{Pinte_ea_2018a}, less significantly inclined disks become a viable option. 

\subsection{Zeeman Observations}
\label{sec:zeeman_observations}

The Zeeman effect, the splitting of lines due to interactions with a magnetic field, can be used to measure the line of sight magnetic field strength \citep[e.g.][]{Vlemmings_ea_2010, Crutcher_2012}. The magnetic field structure in a protoplanetary disk is widely believed to consist of a global poloidal field (such that the magnetic field is along the line-of-sight for a face-on disk), dragged into a toroidal field (perpendicular to the light-of-sight for a face-on disk) by the rotation of the disk in denser regions \citep[e.g.][]{Flock_ea_2015}. As the level of Zeeman splitting is proportional to the line-of-sight magnetic field strength, it is essential to understand where CN is tracing such that one can distinguish between strong toroidal magnetic fields or weak poloidal magnetic fields for face-on disks.

Recently, \citet{Vlemmings_ea_2019} reported tight upper limits for the vertical magnetic field strength $< 0.8$~mG and the toroidal field strength of $< 30$~mG in TW~Hya based on the non-detection of circularly polarized CN emission in TW~Hya. Simulations suggest that typical toroidal fields are on the order of 10 -- 15~mG \citep{Bai_2015}, consistent with the reported upper limits. Assuming that CN is tracing a region in the disk where the magnetic field structure is toroidal, the results in this paper suggest that the densities at $z\,/\,r \sim 0.2$ are sufficient to generate such a morphology. These constraints, in addition to additional tracers of the local gas density and searches for polarized molecular line emission, will place some of the most stringent constraints on the magnetic field morphology in TW~Hya.

\subsection{Future Direction}
\label{sec:discussion:future_direction}

Spectrally resolved observations of CN emission hold a tremendous amount of information on the local conditions of the disk, in particular the gas temperature and collider density. To fully exploit such spectra, an improvement in the modeling approach is needed. While this is a task which is beyond the scope of this paper, we discuss briefly areas which need particular work.

Although radiative transfer codes like \texttt{RADEX} are able to consider non-LTE effects, they do not consider the effects of blending between transitions. This must be taken into account for both the calculation of the level populations and in the subsequent ray-tracing. The latter part is exacerbated by the complex geometries of most protoplanetary disks which are at least moderately inclined, unlike TW~Hya, where high spatial resolution observations would be able to resolve the warm molecular layer above and below the midplane \citep[HD~163296, for example;][]{Rosenfeld_ea_2013}.

In a similar vein, gradients in the gas temperature and density along the line of sight must be included. As the LTE modeling has shown, the optical depths for each hyperfine component within a fine-structure group can vary by an order of magnitude. In this regime it is highly likely that the different hyperfine components are tracing different vertical extents and thus physical properties along the line of sight. Accounting for these gradients in the modeling would potentially allow, in combination with the high-spectral resolution data possible with the stacking technique discussed in Section~\ref{sec:observations:blended_components}, one to place constraints on the line of sight temperature gradient.

We note that there are radiative transfer codes available which provide much of this functionality, for example \texttt{LIME} \citep{Brinch_Hogerheijde_2010}. However, these codes are designed to model entire disks (or any large 3D object), requiring additional assumptions to be made about the underlying disk structure in order to produce the azimuthally averaged spectra we are interested in \citep[for example][]{Guzman_ea_2015}. While these codes can be re-purposed to consider slab-like models, the time required to run such calculations are prohibitively long for the fitting procedure described in Section~\ref{sec:discussion:nonLTE}. Adapting these codes to calculate single spectra would make such fitting problems significantly more tractable and will be a focus of future work.

\section{Summary}
\label{sec:summary}

We have presented a suite of observations of CN emission in the disk around TW~Hya. Out of the 30 hyperfine transitions covered in the set up, 24 are detected. All emission shares a similar radial morphology to previously detected simple molecular species in TW~Hya characterized by a bright ring peaking at 45~au and a shelf of low level emission extending out to $\approx 200$~au.

By accounting for the disk rotation, individual spectra were shifted back to a common velocity before being stacked in concentric annuli. This approach allowed for a resampling of the spectral resolution revealing slight deviations from a Gaussian line profile for two blended components. By fitting two Gaussian components to these transitions it was possible to measure a splitting between these components which were previously unresolved from laboratory measurements.

Calculation of the excitation temperature using a LTE-assumption found that $T_{\rm ex} \gtrsim T_{\rm kin}$ for nearly all fine structure groups, with the difference being largest for the higher frequency $N=3-2$ transitions. Rather than suggesting super-thermal excitation, we interpret this as the manifestation of non-LTE effects in a LTE modeling framework. Using the 0-D radiative transfer code \texttt{RADEX} we show that non-LTE effects are present for CN emission when the local density drops to $\lesssim 10^{-7}~{\rm cm^{-3}}$.

Comparison of the disk average kinetic temperature, $T_{\rm kin} = 24 \pm 5$~K, and a local collider density of $\lesssim 10^{-7}~{\rm cm^{-3}}$ were consistent with the physical conditions at $z\,/\,r \gtrsim 0.2$ in the TW~Hya physical structure model of \citet{Qi_ea_2013}. This is consistent with the formation via vibrationally excited H$_2$ which \citet{Cazzoletti_ea_2018} showed naturally produced the ringed emission morphology seen in many disks.

\acknowledgments
The authors thank the anonymous referee for a constructive report. RT would like to thank St\'{e}phane Guilloteau, Viviana Guzm\'{a}n and Pierre Hily-Blant for helpful discussion on the modeling of CN line emission. RT acknowledges support from the Smithsonian Institution as a Submillimeter Array (SMA) Fellow. This paper makes use of the following ALMA data: 2013.1.00387.S, 2016.1.00440.S and 2017.1.01199.S. ALMA is a partnership of ESO (representing its member states), NSF (USA) and NINS (Japan), together with NRC (Canada), NSC and ASIAA (Taiwan), and KASI (Republic of Korea), in cooperation with the Republic of Chile. The Joint ALMA Observatory is operated by ESO, AUI/NRAO and NAOJ.

\vspace{5mm}
\facilities{ALMA}
\software{
    \texttt{astropy} \citep{astropy_2013, astropy_2018},
    \texttt{emcee} \citep{emcee},
    \texttt{GoFish} \citep{GoFish},
    \texttt{matplotlib} \citep{Hunter_2007},
    \texttt{radex} \citep{vanderTak_ea_2007},
    \texttt{scipy} \citep{scipy}.
}

\bibliography{bibliography}{}
\bibliographystyle{aasjournal}

\appendix

\section{Measuring the Excitation Temperature}
\label{sec:app:Tex}

In this Appendix we detail the procedure followed to measure $T_{\rm ex}$ for each of the fine structure groups, that is a group of transitions which share the same $J$ quantum numbers. This follows the approach presented in the Appendix of \citet{Hily-Blant_ea_2013}, which has been previously used to model CN emission in protoplanetary disks \citep[e.g.][]{Teague_ea_2016, Hily-Blant_ea_2017}.

\subsection{Method}
\label{sec:app:Tex:method}

We assume that each fine structure group shares a common excitation temperature, $T_{\rm ex}$ and that within each fine structure group, the optical depth of each individual transition, $\tau_i$ is assumed to be a fraction of the total optical depth of the fine structure group, $\tau_{i} = r_i \tau_0$, where $r_i$ is the relative intensity of the transition given by,

\begin{equation}
    r_i = \frac{g_{u,\, i} A_{ul,\,i}}{\sum_j g_{u,\,j} A_{ul,\,j}}
\end{equation}

\noindent where $A_{ul}$ is the Einstein-A coefficient of the transition and $g_u$ is the weight of the upper energy level, and $j$ sums over all transitions in the fine structure group.

Assuming that the intrinsic line profile is a Gaussian, i.e. the profile is dominated by Doppler broadening, then the total optical depth can be given by,

\begin{equation}
    \tau(v) = \tau_0 \times \sum_i r_i \exp \left( -\frac{(v - v_{0,\,i})^2}{\Delta V^2} \right)
    \label{eq:app:tau}
\end{equation}

\noindent where $\Delta V$, the line width, is assumed to be constant among all hyperfine components and $v_{0,\, i}$ is the line center of the $i$th transition. We note that we do not provide any special treatment for the blended components and assume that the total optical depth is sufficiently small that it can be summed together.

To convert this optical depth profile to a measure of the flux, we use,

\begin{equation}
    I_{\nu}(v) = \big( B_{\nu}(T_{\rm ex}) - B_{\nu}(T_{\rm bg}) \big) \cdot \big(1 - \exp(-\tau(v)) \big)
    \label{eq:app:Inu}
\end{equation}

\noindent where $B_{\nu}$ is the Planck function, and $T_{\rm bg}$ is the cosmic microwave background temperature. This intensity is converted to a flux density unit through $S_{\nu} = I_{\nu} / \Omega$, where the beam area, $\Omega$, is measured in steradians. It should be noted that an equivalent transformation, using the radiation temperature, $J_{\nu}$, rather than the full Planck function, $B_{\nu}$, will recover the brightness temperature, $T_b$. Thus, the emission for a fine structure group can be modeled with the parameters $\{T_{\rm ex},\, \Delta V, \tau_0\}$ if the line centers are known.

\begin{figure*}
    \centering
    \includegraphics{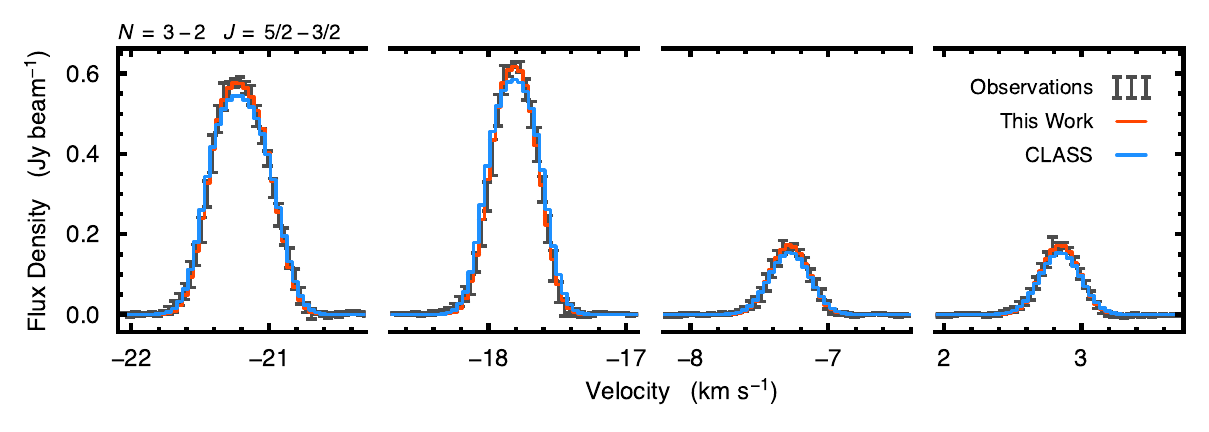}
    \caption{Comparing the fits using the method used in this paper, red, and those using the \texttt{CLASS} software, blue for a spectrum taken at 1\arcsec{}. While both reproduce the observations well, the use of the full Planck law allows for a better fit to the data.}
    \label{fig:app:CLASS_comparison}
\end{figure*}

We note that this procedure is slightly difference from the one in the \texttt{GILDAS/CLASS} software\footnote{\url{http://www.iram.fr/IRAMFR/GILDAS}} which uses the Rayleigh-Jeans approximation for the Planck function, $B_{\nu,\,{\rm RJ}} = 2 \nu^2 k T \, / \, c^2$. At an excitation temperature of 24~K, using the Rayleigh-Jeans approximation causes an underestimation of $I_{\nu}$ by a factor of 0.79 at 340~GHz, 0.88 at 226~GHz and 0.96 at 113~GHz, resulting in differences in $T_{\rm ex}$ of $\sim 5$~K and a factor of $\sim 3$ in $\tau$ for the best-fit values. A comparison of the two fitting methods is shown in Fig.~\ref{fig:app:CLASS_comparison}.

\subsection{Fitting}
\label{sec:app:Tex:fitting}

\begin{figure*}
    \centering
    \includegraphics{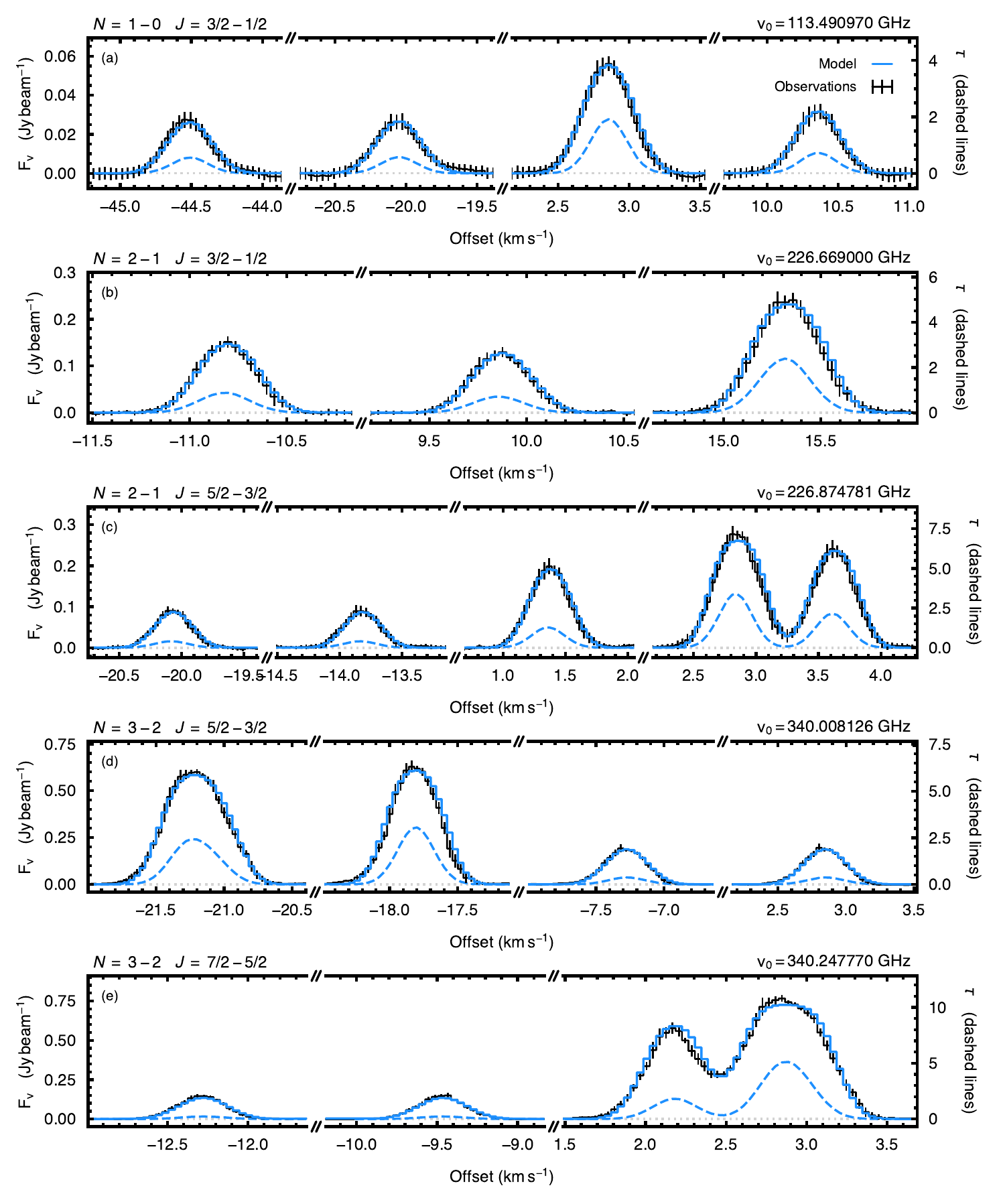}
    \caption{Examples of the best-fit spectra for the five fine-structure groups labeled at the top left of each panel. The black line shows the azimuthally averaged and stacked spectra and the associated $1\sigma$ uncertainties. The solid and dashed blue lines show the median flux density and optical depth of 150 samples of the posterior distributions, respectively. The frequency in the top right of each panel shows the rest frequency of each spectral window.}
    \label{fig:app:LTE_spectra}
\end{figure*}

For an accurate comparison between bands, the data were convolved down to a common beam size, guided by the largest semi-major and semi-minor axes of all the images. This was achieved with the \texttt{imsmooth} task in \texttt{CASA} with a target resolution of $0.6\arcsec \times 0.5\arcsec$ at a position angle of $60\degr$. From these images the aligned and stacked spectra were sampled at spectral resolution of $40~{\rm m\,s^{-1}}$.

Although the line profiles were sampled at a higher rate than images were made at, this does not remove systematic effects intrinsic to the line profile such as the spectral response functions for data taken originally at a low spectral resolution \citep[e.g.][]{Koch_ea_2018}. Due to this, we include an additional convolution step to our model, prior to calculating the likelihood. Following \citet{Loomis_ea_2018a} we adopt a Hanning function for the Band 3 data which was imaged without any spectral averaging, while for the Band 6 and 7 data, which were imaged with a spectral averaging of 8 and 16, respectively, we use a boxcar function with a width equal to the native channel width.

In the modeling, we only consider the total line width, $\Delta V$, with the distinction between thermal, non-thermal and systematic broadening processes (such as those involved with the data acquisition and imaging) discussed in Section~\ref{sec:discussion}. To aid in convergence for low signal-to-noise spectra, particularly in the outer disk, we impose a Gaussian prior on $T_{\rm ex}$ such that,

\begin{equation}
    p(T_{\rm ex} \,|\, \Delta V) = \frac{1}{\sqrt{2 \pi} \delta \Delta V} \exp \left[-\frac{1}{2\delta ^2} \left( \frac{2kT_{\rm ex}}{\Delta V \mu m_p} - 1 \right) \right]
\end{equation}

\noindent where $\delta = 0.23$, equivalent to previous limits on non-thermal broadening, $\alpha \lesssim 10^{-2}$ \citep{Teague_ea_2016, Flaherty_ea_2018}. Finally, each hyperfine component also has a floating offset, $\delta v_{0,\,i}$ to account for any possible offsets in the provided transition frequencies. We do not fit the weak transitions, those with $r_i < 0.02$, as they are not sufficiently well detected in the data to provide any constraint on $T_{\rm ex}$. This results in a set of $3 + N_{\rm trans}$ free parameters for each fine-structure group.

To explore the posterior distributions for these parameters, we use the Affine-invariant MCMC ensemble sampler in \texttt{emcee} \citep{Foreman-Mackey_ea_2013}. A number of walkers equal to twice the number of free parameters was chosen for each run (3 for the physical properties and an additional parameter describing an offset from the laboratory measured line center, $\delta v_{0,\,i}$, for each transition). Values which maximize the likelihood were used as starting positions for the walkers, found by minimizing $\chi^2$ using the `Nelder-Mead' algorithm in \texttt{scipy.optimize.minimize}. Each walker took 2,000 steps to burn in, then an additional 10,000 to sample the posterior distribution. Figure~\ref{fig:app:LTE_spectra} compares the true spectra extracted from the CN ring peak at 0.85\arcsec{} in black, and the modeled spectra in blue. We do not find any significant deviations in the line center, $\delta v_{0,\, i}$, for any of the lines relative to the line centers quoted in Table~\ref{tab:observations}.

\section{Measuring the Kinetic Temperature}
\label{sec:app:Tkin}

In this Appendix, we describe the approach used to infer the underlying kinetic temperature, $T_{\rm kin}$, building upon the forward modeling approach discussed in \citet{Teague_ea_2016} to disentangle broadening effects due to beam convolution.

\subsection{Method}
\label{sec:app:Tkin:method}

In brief, as the line profiles are assumed to be Doppler broadened, the line width can be used as a proxy of the local gas temperature without incurring and uncertainty due to the flux calibration of the data,

\begin{equation}
T_{\rm kin} \leq \Delta V^2 \frac{m_p \mu_{\rm CN}}{2k},
\label{eq:Tkin}
\end{equation}

\noindent where $\mu_{\rm CN}= 26$, $m_p$ is the mass of a proton and $k$ is Boltzmann's constant. However, as discussed in \citet{Teague_ea_2016, Teague_ea_2018b}, the line can be broadened due to non-thermal broadening (turbulence) and the convolution of the beam, inflating the inferred $T_{\rm kin}$ and hence the inequality in Eqn.~\ref{eq:Tkin}. In TW~Hya, the non-thermal broadening component is believed to be small, with upper limits consistent with an uncertainty on $T_{\rm kin}$ of $\lesssim 1$~K \citep{Hughes_ea_2011, Teague_ea_2016, Teague_ea_2018b, Flaherty_ea_2018}. Due to this, we ignore any non-thermal contribution to the line width and focus on the more significant effect of the systematic broadening by beam convolution.

To estimate the systematic broadening associated with the beam convolution we follow a similar forward modeling approach as used in \citet{Teague_ea_2016}. In brief, a model with a known underlying line width profile is convolved with a beam equal to the synthesized beams of each 
observation and resampled to the same spectral resolution as the images. This model observation is subject to the post-processing described in Section~\ref{sec:app:Tex:fitting}. That is, the model observation is first deprojected into concentric annuli, each of which is further spectrally deprojected before being stacked and resampled at $40~{\rm m\,s^{-1}}$ velocity resolution. This process yields an average spectrum at each radius to which a Gaussian function is fit to recover the line width as a function of radius. This can then be compared to the observed radial $\Delta V$ profile.

For the above approach we considered a simple, geometrically thin model comprised of Gaussian emission lines characterized by a width $\Delta V$, peak, $T_B$ and line center, $v_0$. The line center was given by the projected Keplerian rotation around a star with $M_{\rm star} = 0.65~M_{\rm sun}$ and an inclination of $i = 6.8\degr$, i.e. the same used as to deproject the observations. The width was assumed to be entirely thermally broadened such that

\begin{equation}
    \Delta V(r) = \sqrt{2kT_{\rm kin}(r) \, / \, \mu m_p}
    \label{eq:app:dV}
\end{equation}

\noindent where the temperature was parameterised as a broken power law profile of the form

\begin{equation}
T_{\rm kin}(r) =
\begin{cases}
    T_{\rm kin,\,0} \times (r / 1\arcsec\,)^{q_1} \quad \text{where} \quad r < 1\arcsec\\
    T_{\rm kin,\,0} \times (r / 1\arcsec\,)^{q_2} \quad \text{where} \quad r \geq 1\arcsec
\end{cases}
,
\end{equation}

\noindent with $T_{\rm kin,\,0}$ describing the temperature at 1\arcsec{} (60~au) and $q_1$ and $q_2$ the radial dependence. A single power law profile was initially adopted but was found to be too inflexible to to match the data. The choice of 1\arcsec{} as the pivot point is somewhat arbitrary, but reflects the need to consider an `inner' and `outer' disk component. The large beam sizes of $\sim 0.6\arcsec$ essential smooth over this discontinuity such that the location of the pivot point does not substantially affect the results. Here, a positive $q_i$ value does not imply that the bulk temperature profile of the disk is increasing with radius. Rather, as $T_{\rm kin}$ describes the temperature at the height of the CN emission which can change in height as a function of radius, a positive $q_i$ describes a situation where the CN emission height increase with radius, thus tracing warmer regions of the disk. Although the absolute intensity value is not important for the model, the \emph{shape} of the adopted intensity profile will influence the level of broadening, particularly in regions where there is a large gradient \citep[see the Appendix of][for example]{Keppler_ea_2019}. To best match the data, the average of the intensity profiles shown in Fig.~\ref{fig:radial_profiles} was modeled as the sum of three Gaussian components.

\begin{figure}
    \centering
    \includegraphics[]{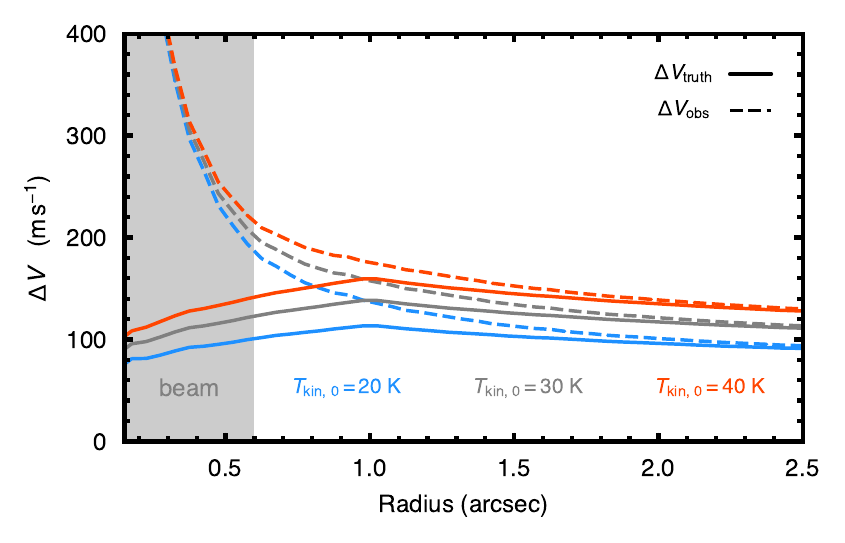}
    \caption{Comparing the intrinsic line width in solid lines and the measured line widths in dashed lines. The three colors represent three different $T_{\rm kin,\,0}$ values. This highlights that even if the molecular region traces a cooler region in the inner disk, the beam convolution will result in a large measured line width.}
    \label{fig:dV_broadening}
\end{figure}

Thus, with a choice of $\{T_{\rm kin,\,0},\, q_1,\, q_2\}$ a model of the observed $\Delta V$ profile can be produced. Figure~\ref{fig:dV_broadening} shows the effect of this systematic broadening of the line using a model with a range of $T_{\rm kin,0}$ values with $q_1 = 0.5$ and $q_2 = -0.5$. As discussed in \citet{Teague_ea_2016}, this broadening is more prominent in the inner disk at radii within two beam major axes. In the outer disk where the velocity gradient across the beam is minimal, the additional broadening is significantly reduced.

\subsection{Modeling}
\label{sec:Tkin:modeling}

To match the data, the line widths from the $T_{\rm ex}$ fitting was used. As an addition check, each individual component was fit with a Gaussian profile to verify that the $\Delta V$ values from Section~\ref{sec:app:Tex:fitting} were not strongly affected by blended components. Due to the blending of the strong transitions and the marginal detection of the weaker components, we chose not to fit for the $J = 7/2 - 5/2$ and $J = 3/2 - 1/2$ fine structure groups of the $N = 3 - 2$ transition. Given the low signal level in the outer disk, only the widths between 0.3\arcsec{} and 3.0\arcsec{} were considered for the fit, apart from the $N = 3-2$, J$=5/2-3/2$ fine structure group which was only fit out to 2.0\arcsec{} due to the blending of the hyperfine components resulting in over-estimated $\Delta V$ profiles. As before, \texttt{emcee} \citep{emcee} was used to sample the posterior distributions of $\{T_{\rm kin,\,0},\, q_1,\, q_2\}$. 12 walkers were used for each fit, starting at positions which were found to maximize the likelihood (found using the \texttt{curve\_fit} function within \texttt{scipy.optimize}). Walkers took 2,000 steps to burn in and an additional 2,000 steps to sample the posterior distributions.

\begin{deluxetable}{ccccc}
\tablecaption{Inferred Kinetic Temperature Profiles \label{tab:tkin}}
\tablehead{
\colhead{$N^{\prime} - N^{\prime\prime}$} &
\colhead{$J^{\prime} - J^{\prime\prime}$} &
\colhead{$T_{\rm kin,\,0}$} &
\colhead{$q_1$} &
\colhead{$q_2$}
}
\startdata
$1 - 0$ & $3/2 - 1/2$ & $28 \pm 3$~K & $+0.2 \pm 0.5$ & $-0.6 \pm 0.3$ \\
$2 - 1$ & $3/2 - 1/2$ & $31 \pm 3$~K & $+0.4 \pm 0.5$ & $-0.6 \pm 0.2$ \\
$2 - 1$ & $5/2 - 3/2$ & $29 \pm 2$~K & $-0.1 \pm 0.3$ & $-0.5 \pm 0.1$ \\
$3 - 2$ & $5/2 - 3/2$ & $29 \pm 2$~K & $-0.1 \pm 0.2$ & $-0.1 \pm 0.1$ \\
\hline
\enddata
\tablecomments{Uncertainties represent the 16th to 84th percentiles of the posterior distributions which were found to be symmetric about the median value.}
\end{deluxetable}

Table~\ref{tab:tkin} describes the resulting posterior distributions of the parameters for the four fine structure groups. Figure~\ref{fig:TexTkin} compares the inferred $T_{\rm kin}$ radial profiles with the measured LTE $T_{\rm ex}$ profiles from Section~\ref{sec:app:Tex:fitting}.

\end{document}